\documentclass[]{spie}  

\usepackage{amsmath,amsfonts,amssymb}
\usepackage{graphicx}
\usepackage[colorlinks=true, allcolors=blue]{hyperref}

\title{Characterization of a Longwave HgCdTe GeoSnap Detector}

\author[a]{Rory Bowens}
\author[a]{Michael R. Meyer}
\author[a]{Taylor L. Tobin}
\author[a]{Eric Viges}
\author[a]{Dennis Hart}
\author[a]{John Monnier}
\author[b]{Jarron Leisenring}
\author[c]{Derek Ives}
\author[d]{Roy van Boekel}
\affil[a]{The University of Michigan, 500 S State Street, Ann Arbor, USA}
\affil[b]{The University of Arizona, 1200 E University Boulevard, Tucson, USA}
\affil[c]{European Southern Observatory, Karl-Schwarzschild-Straße 2, Garching bei München, Germany}
\affil[d]{Max Planck Institute for Astronomy, Königstuhl 17, Heidelberg, Germany}

\authorinfo{Primary Author Email: rpbowens@umich.edu}

\begin{document} 
\maketitle

\begin{abstract}
New longwave HgCdTe detectors are critical to upcoming plans for ground-based infrared astronomy. These detectors, with fast-readouts and deep well-depths, will be key components of extremely large telescope instruments and therefore must be well understood prior to deployment.
We analyze one such HgCdTe detector, a Teledyne Imaging Sensors GeoSnap, at the University of Michigan.
We find that the properties of the GeoSnap are consistent with expectations from analysis of past devices. The GeoSnap has a well-depth of 2.75 million electrons per pixel, a read noise of 360 e-/pix, and a dark current of 330,000 e-/s/pix at 45 K. The device experiences 1/f noise which can be mitigated relative to half-well shot noise with modest frequency image differencing. The GeoSnap's quantum efficiency is calculated to be 79.7 $\pm$ 8.3 \% at 10.6 microns. Although the GeoSnap's bad pixel fraction, on the order of 3\%, is consistent with other GeoSnap devices, close to a third of the bad pixels in this detector are clustered in a series of 31 "leopard" spots spread across the detector plane.
We report these properties and identify additional analyses that will be performed on future GeoSnap detectors.

\end{abstract}


\section{INTRODUCTION}
\label{sec:intro}
Ground-based astronomy will be revolutionized by extremely large telescopes, such as the European Southern Observatory's Extremely Large Telescope (henceforth ELT). These facilities offer unprecedented spatial resolution and light gathering power that will be utilized by a powerful suite of instruments. The Mid-infrared ELT Imager and Spectrograph (METIS) will be the ELT's window into the 3-13 micron regime \cite{brandl2021Msngr.182...22B}. METIS will be able to probe a range of science topics including protoplanetary disks, forming planets, atmospheres of mature planets, young stellar clusters, evolved stars, the galactic center, and active galactic nuclei. This includes the potential for a first direct image of a small, temperate exoplanet around one of the nearest stars \cite{bowens2021EPSC...15...79B}.

METIS will utilize a GeoSnap detector with a B0 readout integrated circuit (ROIC), developed by Teledyne Imaging Sensors (TIS). GeoSnap is a 2048x2048 HgCdTe (MCT) detector with wavelength range from $<$ 3 to $>$ 13 microns, high quantum efficiency (QE), and favorable noise properties compared to existing longwave Si:As blocked-impurity-band detectors\cite{mcmurtry2013OptEn..52i1804M}. The ROIC utilizes a capacitance transimpedance amplifier (CTIA). CTIAs are useful for mid-IR applications as they can achieve high well depths (order of 2 million electrons) and should not suffer from persistence\cite{smith2008SPIE.7021E..0JS} which is present in the source follower method employed in other state-of-the-art detectors such as the HAWAII-2RG\cite{beletic2008SPIE.7021E..0HB}. GeoSnap can operate at 86.7 Hz in standard operational mode (and up to 143.2 Hz with certain settings) and has an 18 micron pixel pitch. At the University of Michigan (UM), we have begun testing potential candidate GeoSnap detectors for use in METIS, starting with the GeoSnap SN23260. SN23260 represents the first of at least two candidate detectors that we are testing for the METIS/ESO team. We focused on testing core properties of SN23260 including its cosmetics, gain, and dark current. We compared SN23260 with an A0 ROIC GeoSnap detector (henceforth UM GeoSnap) previously characterized by the UM and the University of Arizona (UA)\cite{jarron2023AN....34430103L}. The UM GeoSnap has only one of its four quadrants hybridized with MCT material. The MCT material originated from long wavelength development programs at the University of Rochester \cite{cabrera2019JATIS...5c6005C}. We anticipate a different dark current owing to a different long wavelength cutoff (see the empirical Rule 07 \cite{tennant2008JEMat..37.1406T}) and a different read noise owing to different well depths between the two instruments.

We begin the paper with a description of the testing conditions at UM (Section \ref{sec:testingcond}). We then describe the results for SN23260 (Section \ref{sec:testingresults}), covering the cosmetics, gain, well-depth, read noise, bias, dark current, linearity, 1/f noise, and quantum efficiency. Finally, we summarize our results and suggest future work (Section \ref{sec:conclusion}).

\section{Testing Conditions}
\label{sec:testingcond}

\subsection{MITTEN Cryostat}

We tested SN23260 in the Michigan Infrared Test Thermal ELT N-band (MITTEN) Cryostat at UM. Details of MITTEN were reported in Bowens et al. 2020\cite{bowens2020SPIE11447E..37B} but we briefly summarize its major features and new additions here. MITTEN's interior volume is a cylinder that is 36.3 cm tall and 67 cm in diameter. The optical bench and surrounding shield reach temperatures $<$ 20 K while other interior surfaces reach temperatures $<$ 60 K. The system has two thermal shrouds and no windows and is cooled via a pulse-tube cryo-cooler.

\begin{figure}
    \centering
    \includegraphics[width=0.8\textwidth]{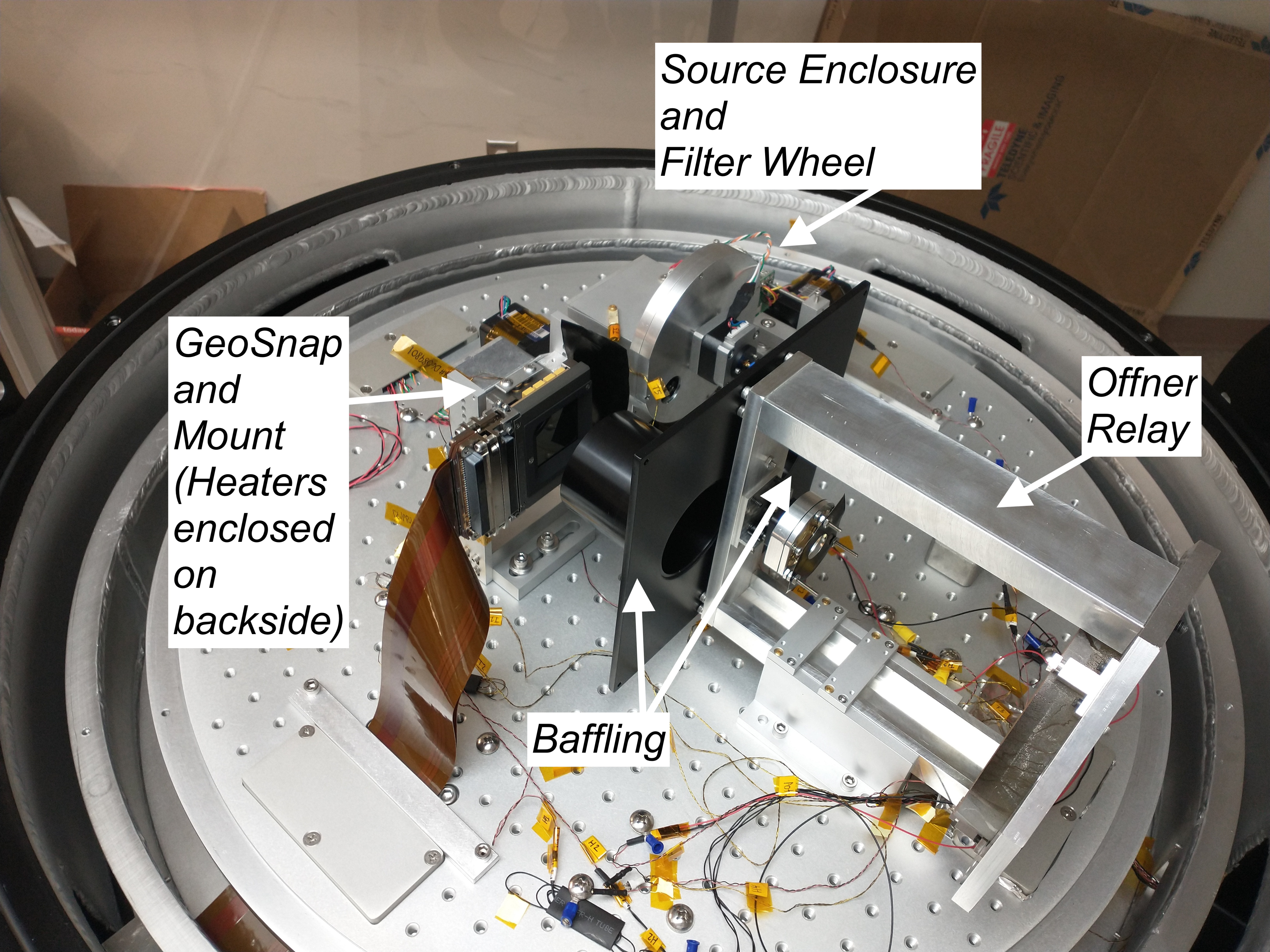}
    \includegraphics[width=0.4\textwidth]{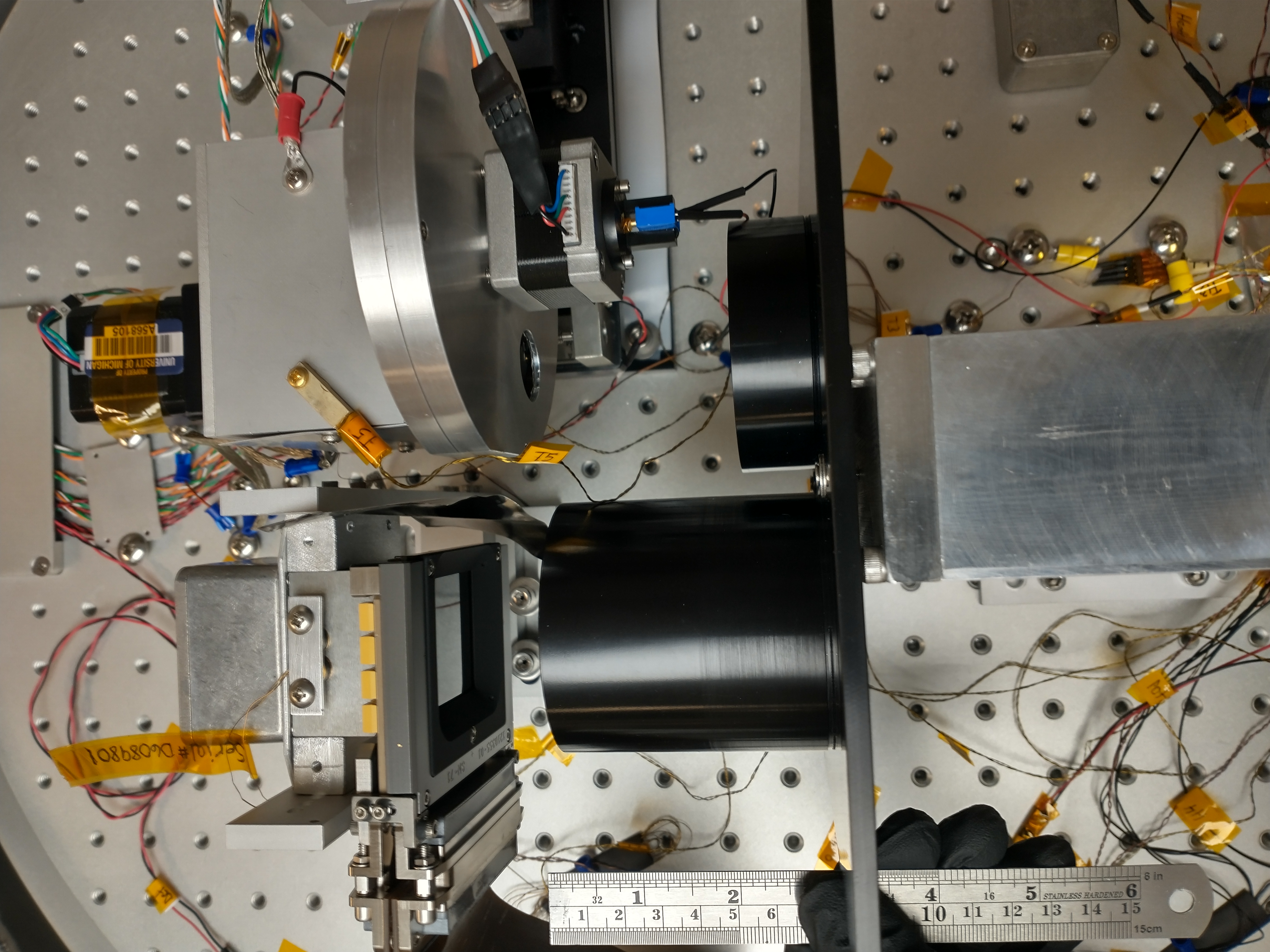}
    \includegraphics[width=0.4\textwidth]{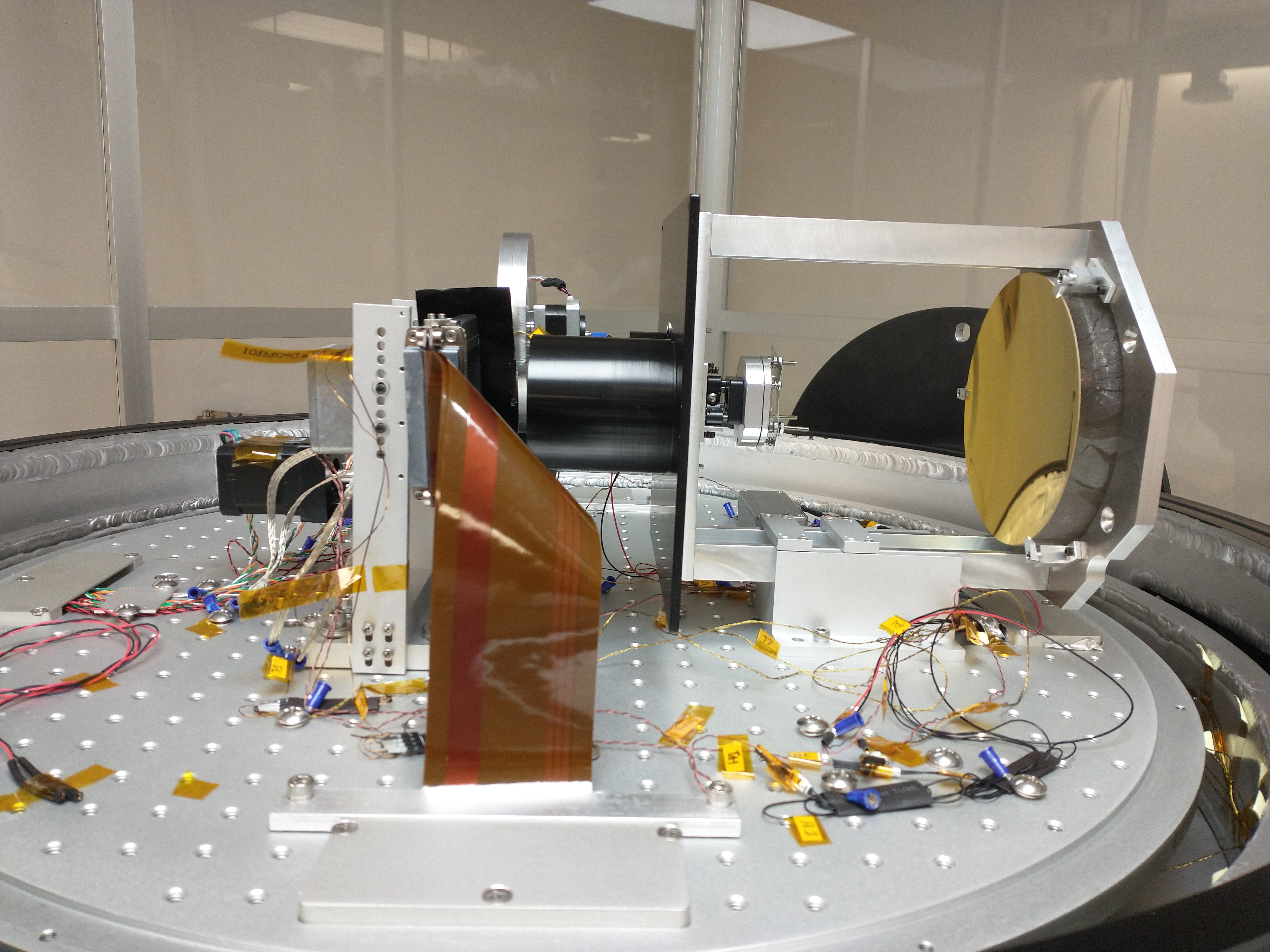}
    \caption{Photographs of the MITTEN Optical Bench. The layout of the optical bench is similar to the description in Bowens et al. 2020 \cite{bowens2020SPIE11447E..37B} with two important changes. First, the baffling on the Offner relay is shorter in length but larger in diameter, in an effort to illuminate the now four times larger detector surface. Second, the heaters behind the cryostat mount are now encased in an aluminum enclosure to reduce thermal contamination.}
    \label{fig:mittenbench}
\end{figure}

A photograph of the interior layout can be seen in Figure \ref{fig:mittenbench}. The testing environment uses 1:1 reimaging via an $f/11$ Offner relay to fit the detector and thermal source within the compact interior space. The detector mount is a molybdenum block whose temperature is controlled by two heaters, encased in 6061 aluminum on the backside of the mount. They can maintain the detector mount to a precision of $\pm$ 0.001 K within the range of 30-70 K. However, the detector introduces approximately 0.8 W of heating (which can vary slightly depending on selected frame rate). In an environment where the user is changing frame rate on the order of minutes (such as during our testing), it is more accurate to say the mount can be maintained to a precision of $\pm$ 0.03 K.

Adjacent to the detector is the light source. The source is a thermally isolated copper block with a centrally located heating element (and offset temperature diode) on the backside. The front-side is coated with ultra-black nano particles to improve the emissivity. During testing, the source was held within $\pm$ 0.03 K of the set-point temperature. The entire copper source block is enclosed in a cold aluminum case. Light exits via a 2.03 cm diameter exit hole.

The exit hole is flush to a slide that can be used to restrict the field of view to a series of small pinholes. Then the light travels through a six position filter wheel which includes one blank position and one open position. These structures are all enclosed except their co-aligned 2.03 cm diameter entrances and exits to reduce stray light. The filter wheel represents the greatest uncertainty in alignment but should maintain a center within 1 mm of source's center. Finally, the light exits the filter wheel enclosure, travels through the first set of baffling, travels through the Offner relay, and then the second set of baffling to a focus on the detector. The co-aligned equal diameter exits at the source result in vignetting but a mask at the secondary mirror is used to define the system as f/11. The source was originally designed for testing the UM GeoSnap used for MIRAC-5 \cite{jarron2023AN....34430103L,bowens2022SPIE12184E..1UB} which only uses one of the four quadrants. As such, it is undersized for the testing of the full 2048x2048 GeoSnaps and can illuminate approximately 25\% of the array. It can be moved laterally across the detector surface but not vertically. Due to a gradient across the source (see Section \ref{subsec:gain}), we choose to leave the source at a fixed position near the center of the detector.  

There are three operational heaters within the cryostat. One attached to the source and two attached to the detector mount. Several diodes are located throughout the cryostat to monitor temperature. Past testing has revealed that even when the source is heated to 300 K, the exterior of the source box does not exceed 40 K. A test filter and the filter wheel were left exposed to a 300 K source and did not exceed 62 K after 6 hours.

Baffling has been increased from 2.9 cm diameter to 6.3 cm diameter to eliminate vignetting of the f/11 beam from the source at its location. The lengths of the baffles had to be decreased due to their larger radii in order to avoid collisions with other interior structures. This results in an approximately two inch gap from the filter wheel enclosure to the first baffle and a one inch gap from the second baffle to the detector surface. A small strip of baffling was inserted on the side of the detector to block potential scattered light from the source.

The filter wheel for MITTEN has also been updated. Descriptions of the new filters are given in Table \ref{tab:filter_table}. One of the six filter positions provides an opaque "blank" position. Another position provides a 2.03 cm diameter "open" position. Only the "ammonia" (named for its scientific goal of measuring the 10.6 micron ammonia feature) and 11 micron filters with their narrow bandpasses and blocking across the whole wavelength range over which the detector is sensitive, are used for QE estimates. Both of these filters display obscuration in a small edge region due to oversized "washers" used to hold them in place. All filters have measured transmission curves from vendors at 77 K\footnote{These filters are expected to have temperatures of 30 to 40 K in the cryostat. Conversations with Materion while acquiring the ammonia filter suggested shifts of about 0.05 microns due to temperature change.} which are used throughout the paper. All transmission curves are from at least 2 to 14 microns, even if the blocking covers a smaller range.

\begin{table}[ht]
\caption{The MITTEN filter wheel has six positions 60 degrees in separation from each other. Each position is 2.03 cm in diameter. Transmission curves are known from at least 2 to 14 microns for all filters.} 
\label{tab:filter_table}
\begin{center}       
\begin{tabular}{|l|l|l|l|l|l|}
\hline
\rule[-1ex]{0pt}{3.5ex}  Filter & Central $\lambda$ & Full Width Half Max & Peak Trans. & Blocking Trans. & Blocking Range\\
\rule[-1ex]{0pt}{3.5ex}  Name & (microns) & (microns) & (\%) & (\%) & (microns)\\
\hline

\rule[-1ex]{0pt}{3.5ex}  Blank & - & - & - & - & -\\
\hline
\rule[-1ex]{0pt}{3.5ex}  Open & - & - & - & - & -\\
\hline
\rule[-1ex]{0pt}{3.5ex}  5 Micron & 5.03 $\pm$ 0.7 & 2.0 $\pm$ 0.7 & $\geq$ 80\% & $\leq$ 1\% & 0.4 to 5.5 \\ 
\hline
\rule[-1ex]{0pt}{3.5ex}  8 Micron & 8.87 $\pm$ 0.14 & 2.23 $\pm$ 0.14 & $\geq$ 80\% & $\leq$ 1\% & 1.5 to 12.0 \\ 
\hline
\rule[-1ex]{0pt}{3.5ex}  Ammonia & 10.59 $\pm$ 0.05 & 0.64 $\pm$ 0.07 & $\geq$ 80\% & $\leq$ 0.1\% & 1.0 to 13.0 \\
\hline
\rule[-1ex]{0pt}{3.5ex}  11 Micron & 11.300 $\pm$ 0.057 & 0.226 $\pm$ 0.045 & $\geq$ 70\% & $\leq$ 0.30\% & 0.300 to $\geq$ 15.000 \\
\hline
\end{tabular}
\end{center}
\end{table}

\subsection{Tuning}
\begin{figure}
    \centering
    \includegraphics[width=\textwidth]{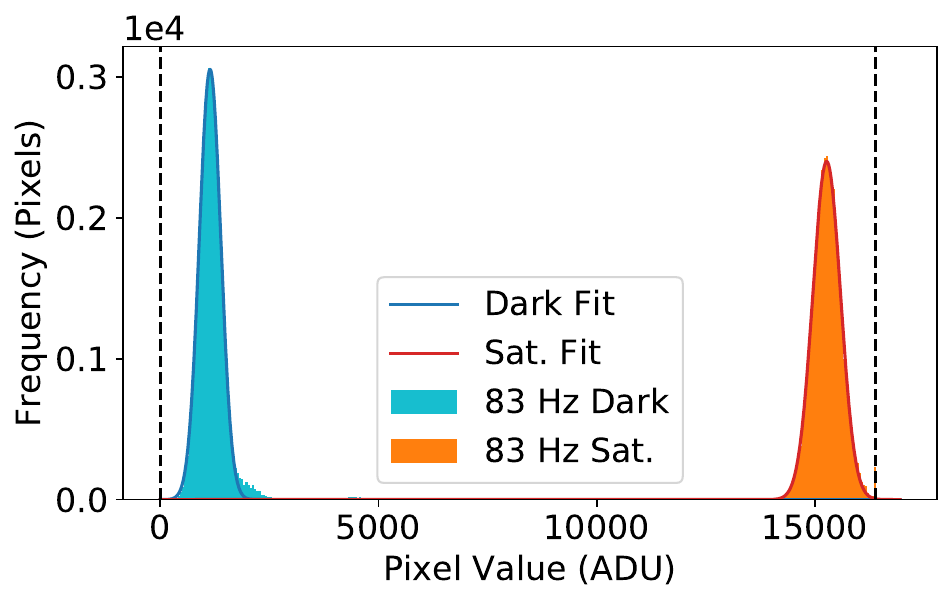}
    \caption{Histograms and fits for a 200 x 200 pixel subregion of a 83 Hz dark (approximately the bias at this frame rate) and 83 Hz saturated image. The dark data is well fit by a Gaussian with mean 1142.0 ADU/pix and sigma 246.5 ADU/pix, over 3-sigma from the 0 threshold. Slight margin is allotted for the difference between the true bias and the 83 Hz dark. Likewise, the saturated data is well fit by a Gaussian with mean of 15270 ADU/pix and sigma of 319 ADU/pix; over 3-sigma from the 16384 threshold. Both thresholds are marked by dashed, vertical lines.}
    \label{fig:tuning_hist}
\end{figure}

We tuned SN23260 at 83 Hz for the MITTEN cryostat following instructions from TIS. SN23260 was held to 45 K $\pm$ 0.03 K. The analog-to-digital converter (ADC) gain and offset were defined such that in a bias image and in a saturated image, a Gaussian fit to the pixel value histogram was 3-sigma from the 0 ADU value and 16,384 ADU value, respectively. This is necessary to ensure that the full dynamic range of the device is utilized without allowing a significant number of pixels to under or overflow their registers. 
We display a histogram of our "bias" (spatial histogram of the mean of ten 83 Hz, 100\% Tint images with the blank filter in front of the 225 K source) and saturated (spatial histogram of the mean of ten 83 Hz, 100\% Tint images of the 225 K source) in Figure \ref{fig:tuning_hist} (see Section \ref{sec:testingresults} for an explanation of Tint). These data utilize a conservative 10 Hz bad pixel mask described in \ref{subsec:cosmetics} and region A (to coincide with a smooth region of the undersized source) described in \ref{subsec:gain}.

\section{Results}
\label{sec:testingresults}

We summarize the series of tests performed on SN23260. Unless otherwise noted, SN23260 is kept at 45 K $\pm$ 0.03 K. Exposures are taken in sets of 128 frames which are then combined via a temporal mean. 

GeoSnap can be read with variable integration time, called Tint. GeoSnap "Tint" is different from the typical use of the word Tint: for GeoSnap it is a percentage of the frame time (which is defined by the frame rate). Thus, a 10 Hz frame with 50\% Tint is only actually exposed for $\sim$0.05 seconds, followed by $\sim$0.05 seconds of dead time with pixels shorted before the next frame begins. We utilize Tint ramps throughout the analysis of SN23260, taking ramps from 0\% to 100\% Tint in 5\% Tint steps. At each step 128 frames are acquired. There are some intricacies to GeoSnap's exposure time with regards to Tint that deserve explanation. The minimum exposure time for GeoSnap is defined by the master clock frequency, typically set to 80 MHz. Tint 0\% actually results in an exposure time of 320 $\mu$s, regardless of user defined imaging frame rate. Likewise, the maximum exposure time is actually the frame time subtracted by the minimum reset time of 140.5 $\mu$s. Thus, Tint 100\% of a 50 Hz frame has a max exposure time of 19.8595 ms, not 20 ms. The actual exposure time at all Tints are slightly different from the simple estimate of Tint multiplied by frame time. At 10 Hz, these differences are on the order of 0.14 ms and are fixed for different Tint/frame rate combos. The process is further complicated by GeoSnap swapping between Integrate then Read (ITR) and Integrate while Read (IWR) modes depending on the frame rate and Tint\footnote{METIS will utilize its own electronics to operate its GeoSnap and therefore these trends are exclusive to our testing environment using the TIS GeoSnap electronics.}. As the device needs enough time to read out the data, it will perform IWR at higher Tint levels where it no longer has enough time to wait in between planned frames. The swap also results in bias offsets. The transition between the modes occurs at $t_{trans} =(t_{max\_exposure\_time} - 11.53\ ms)/t_{max\_exposure\_time}$ where 11.53 ms is the minimum readout time (based on the 86.7 Hz max frame rate) of the detector in 2048x2048 mode. So for a 50 Hz frame rate, the change would be visible at about 43\% Tint.

\subsection{Fraction of Operable Pixels}
\label{subsec:cosmetics}
Before analyzing the data, we produce a bad pixel mask that will be used in all future data analysis. SN23260's cold mask occults pixels close to the detector's edges. Furthermore, Fresnel diffraction occurs directly adjacent to this. In all of our analyses, we exclude the 100 pixels directly adjacent to the edges, effectively limiting our device to a 1848x1848 device. These pixels are not inherently bad but the physical layout of the device prevents their use. We refer to this as the "edge" mask. Furthermore, a horizontal strip due to metadata processing during readout (which will be fixed in future imaging software updates) appears at y = 1024 on the array. These pixels make up 0.05\% of the array and are also excluded from the analysis.

\begin{figure}
    \centering
    \includegraphics[width=0.8\textwidth]{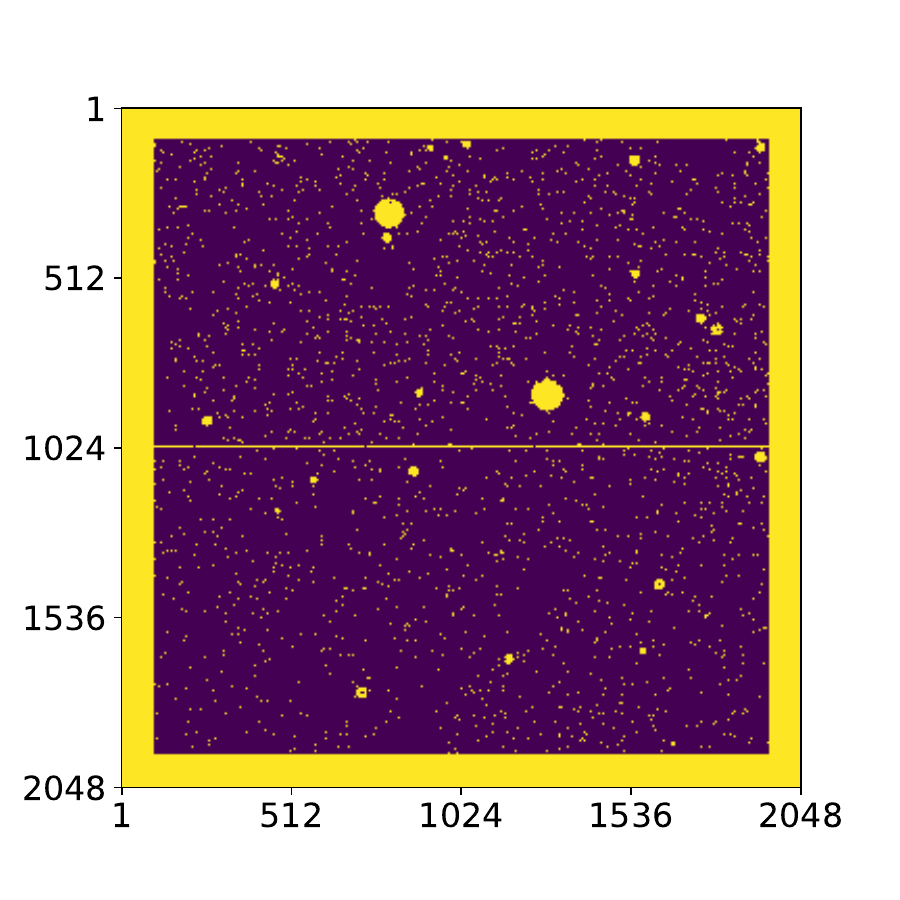}
    \caption{The edge mask and 10 Hz bad pixel mask for SN23260. The two "glow" spots are visible as the two large diameter circles on the mask. A horizontal strip due to metadata processing during readout (which will be fixed in future imaging software updates) appears at y = 1024.}
    \label{fig:badpixel}
\end{figure}

After excluding the edge mask and horizontal metadata strip, we begin making a bad pixel mask, starting with a mean dark frame made from the 128 $\times$ 100\% Tint 10 Hz dark images (i.e., the source cold and the filter wheel in the blank position). This dark is the temporal mean of these 128 dark images, restricted to the inner 1848x1848 pixel region, and based on it we apply two criteria to exclude pixels from analysis. First, we remove pixels with dark currents greater than 20 million e-/s/pix. For 10 Hz data and assuming the gain and bias are approximately 200 e-/ADU and 1000 ADU, respectively, we consider all pixels with values over 11,000 ADU to exceed this dark current limit (hereafter "hot" pixels). We also mask all pixels whose median absolute deviation (MAD) across the 128 frame equals 0. We call these pixels "zero-variance" and they may result from being dead or saturated due to dark current. We find 1.01\% of pixels in the 1848x1848 region are hot and 2.22\% are zero-variance. The criteria are not mutually exclusive so we find that the total number of bad pixels is 3.17\%. The mask for the 10 Hz data is shown in Figure \ref{fig:badpixel}. SN23260 suffers from 29 regions of concentrated bad pixels and 2 regions of excess dark current, which we refer to as "leopard" and "glow" spots, respectively. The two glow spots are regions of excess dark current which decrease in size as exposure time decreases. For consistency, we utilize the 10 Hz mask for all data with frame rates greater than 10 Hz, as this should always cover the two glow spots. In cases where we perform 1 or 5 Hz integrations, we use a 1 or 5 Hz bad pixel mask (constructed in an analogous way), respectively.

 We analyze the leopard and glow spots, focusing on their location and stability for the leopard spots. Analysis of data taken in February 2023 at 85 Hz found the locations of the 29 leopard spots and determined that they could be accurately fit with top-hat functions. We repeat that analysis with our own 85 Hz data from April 2024 generated from a mean of 128 frames. We find that, with the exception of one spot (which is composed of a majority dead pixels), all spots can be fit by top-hat functions with (x,y) positions and radii within 2 pixels of their previously estimated values estimated. The two glow spots likewise remained in the same location between the two data sets and had FWHM 4 pixels larger than their prior values at the same temperature and frame rate. The glow spots are the result of excess dark current and therefore their change in size may be representative of a slightly higher dark current in our testing environment.

Approximately 12,500 pixels are contained within the leopard spots, and, at 10 Hz, 11,000 pixels across both glow spots. This represents approximately 0.7\% of the 1848x1848 working area. The hot and zero-variance pixel percentage excluding the edge mask and the spots then equals approximately 2.5\%.

\subsection{Gain and Well Depth}
\label{subsec:gain}
\begin{figure}
    \centering
    \includegraphics[width=0.63\textwidth]{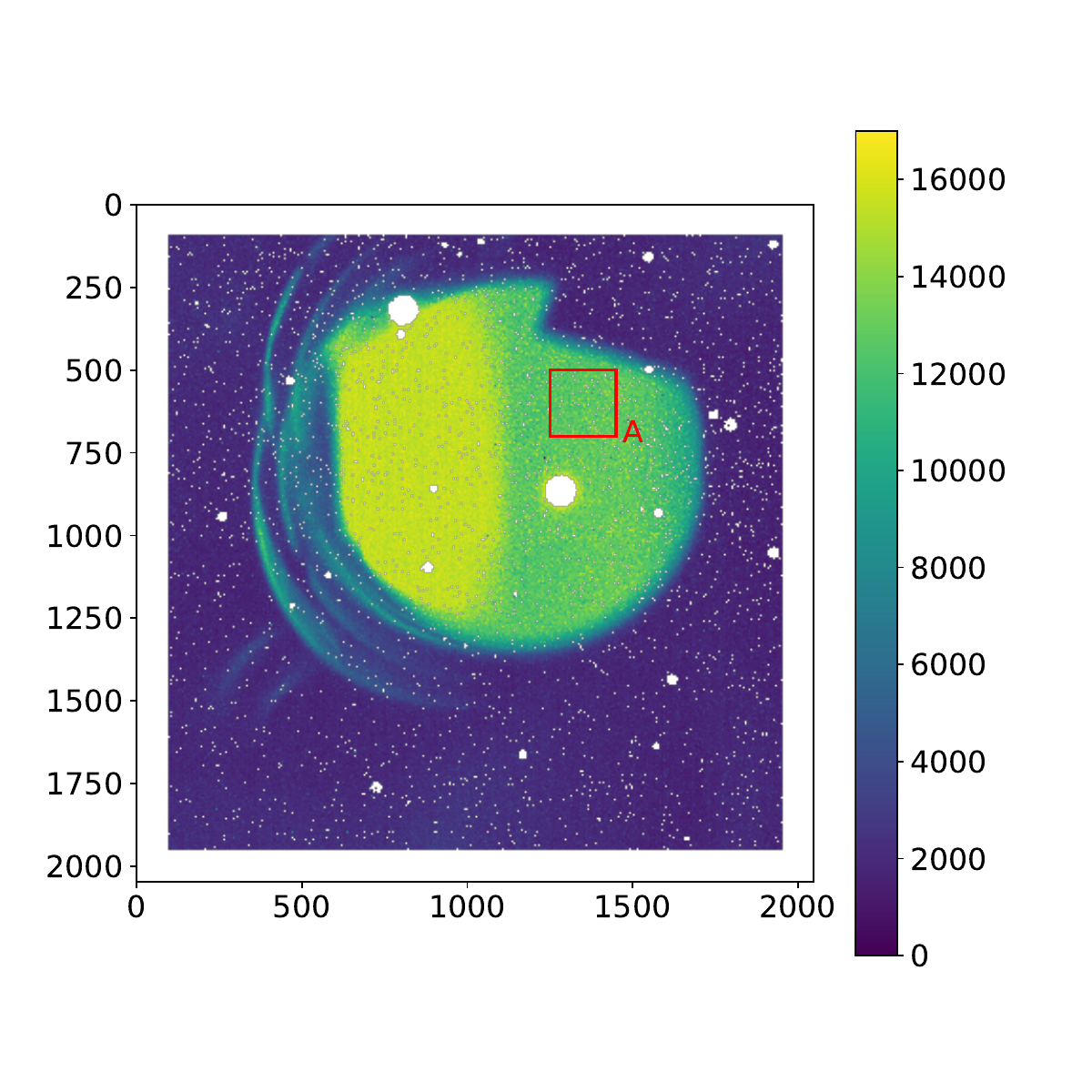}
    \includegraphics[width=0.73\textwidth]{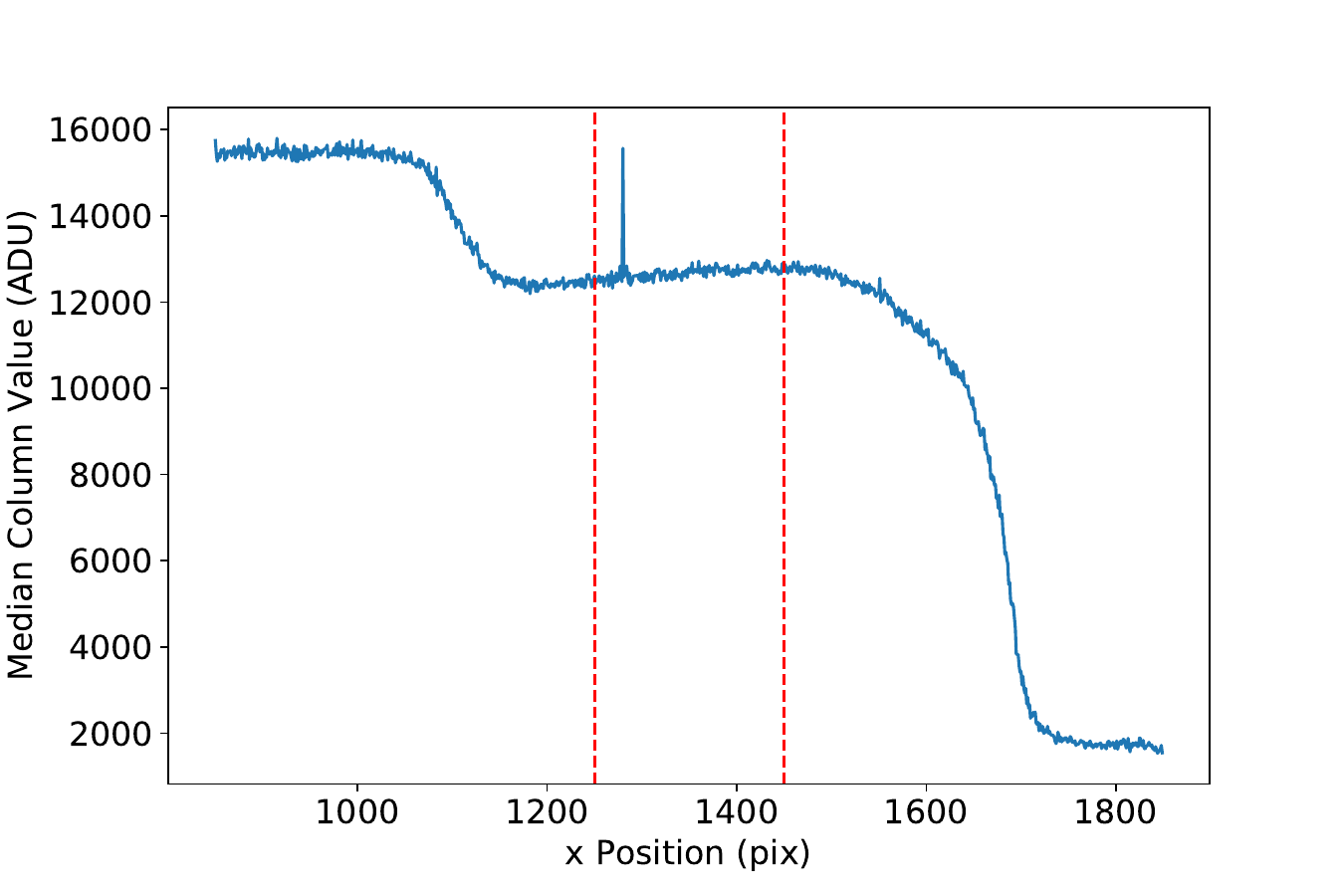}
    \caption{(Top): A picture of the masked, 200 K source at 80\% Tint, 20 Hz, with the ammonia filter in place (including one of its slightly oversized rectangular "washers"). We select a small region of the detector for use in the gain calculation and other tests. (Bottom): The median pixel value within each column for region A and columns adjacent to it. A one-pixel wide spike (not visible in the top picture at this scale) intercepts our region, causing the outlier near pixel 1230.}
    \label{fig:linearity_region}
\end{figure}

To measure the gain, we set the source to 200 K and look through the ammonia filter. We produce ramps in 5\% Tint steps. At each step, we generate a mean frame from the time series of 128 frames, after applying the appropriate bad pixel mask. As noted, the source is undersized for our detector and exhibits a lateral gradient across it which may be due to the aging of the coating. We utilize the region within (1250:1450,500:700) pixels (henceforth region A), as marked in Figure \ref{fig:linearity_region}, for our analysis. The accompanying plot in Figure \ref{fig:linearity_region} shows the median x-value of each column within the region and slightly beyond.

We utilize the region A data to estimate the gain of SN23260. For each Tint step, we calculate a standard deviation per pixel by comparing the values of each pixel across all 128 frames. For each pixel, we perform a linear fit of the pixel's variance versus the median across the first 14 Tint steps (0\% through 60\% Tint). Our gain per pixel is then the inverse of the slope. We select the specific Tint range to avoid any non-linearity as we approach saturation. A Gaussian fit to the gain per pixel histogram returns a mean of 194.7 e-/ADU and sigma of 40.5 e-/ADU (although there are obvious non-Gaussian tails).

We crosscheck our gain result with a different approach that avoids using Tint ramps. We define region B at (750:1500,500:1100), i.e., a rectangular region covering most of the source. We estimate median pixel values and pixel standard deviations across 128 $\times$ 20 Hz and 20\% Tint frames. We solve for the gain per pixel as the median pixel value divided by the pixel variance (excluding bad pixels). We then fit a Gaussian to the histogram, recovering a mean gain of 193.4 e-/ADU and sigma of 31.6 e-/ADU. We use 194.7 e-/ADU gain for the remainder of the paper as it is based on a larger data volume.

Utilizing the fit to the saturated data from Figure \ref{fig:tuning_hist}, we can also estimate the detector's well depth. For a gain of 194.7 e-/ADU and mean saturated pixel value of 14,130 ADU/pix (after removing the offset), our detector has a mean well depth of 2.75 million electrons per pixel.

\subsection{Read Noise}
\begin{figure}
    \centering
    \includegraphics[width=\textwidth]{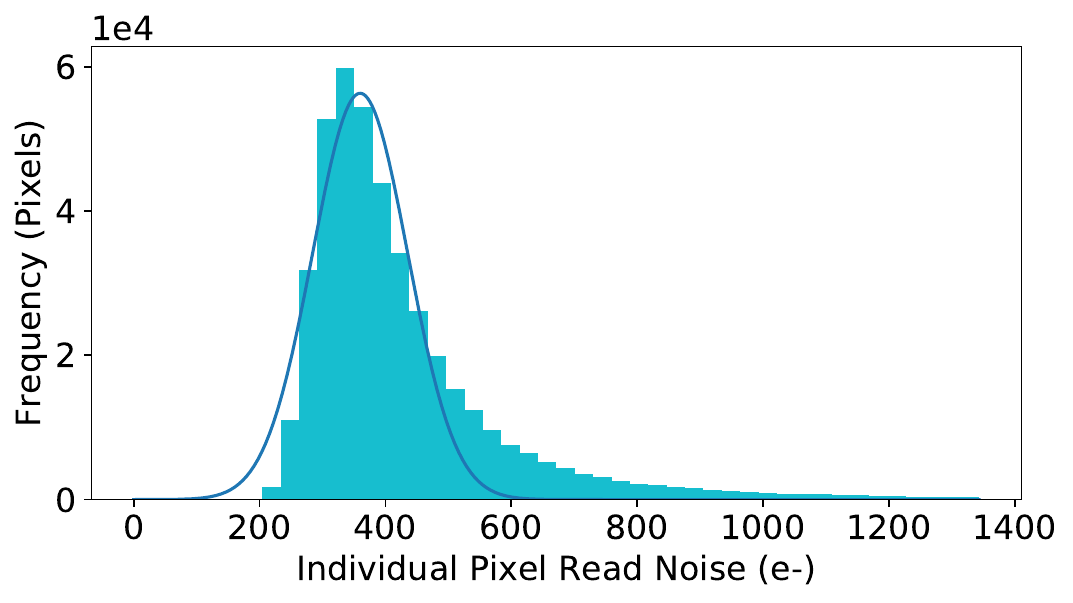}
    \caption{A histogram of pixel noise across 64 pair subtracted frames at 85 Hz, 0\% Tint.}
    \label{fig:readnoise}
\end{figure}

To determine the read noise, we take 128 frames at 85 Hz, 0\% Tint with the source off (T $\sim$ 45 K). We subtract adjacent frames, i.e., frame 1 by frame 2 and then frame 3 by frame 4. We then calculated a standard deviation for each pixel across its 64 values for all 1848x1848 pixels. We plot a histogram of standard deviations, divided by square-root of two (such that they represent the read noise of a single, non-subtracted frame) in Figure \ref{fig:readnoise}. We fit a Gaussian to the data, recovering a mean read noise of 360 e-/pix (1.85 ADU/pix) with sigma of 78 e-/pix (0.40 ADU/pix). The expected shot noise due to dark current at 0.320 ms (the 0\% Tint minimum exposure time) is 16 e-/pix (0.2\% of the total). We also repeat the analysis with 100\% Tint frames at different frame rates. Accounting for shot noise, we find that read noise does vary depending on the frame rate and Tint combination utilized. At 10 Hz and 100\% Tint, we estimate a read noise of 485 e-/pix (2.49 DN/pix) via this method. The read noise appears to be stable for a given Tint and frame rate.

\subsection{Bias and Dark Current}
We estimate the bias and dark current for SN23260. For bias, we assume that all signal contributions at 0\% Tint are negligible (true exposure time is 0.320 ms). We create a mean frame from 128 $\times$ 0\% Tint frames at 1 Hz. We fit a Gaussian to the bias signal per pixel histogram and recover a mean of 1017.0 $\pm$ 0.1 ADU/pix.

\begin{figure}
    \centering
    \includegraphics{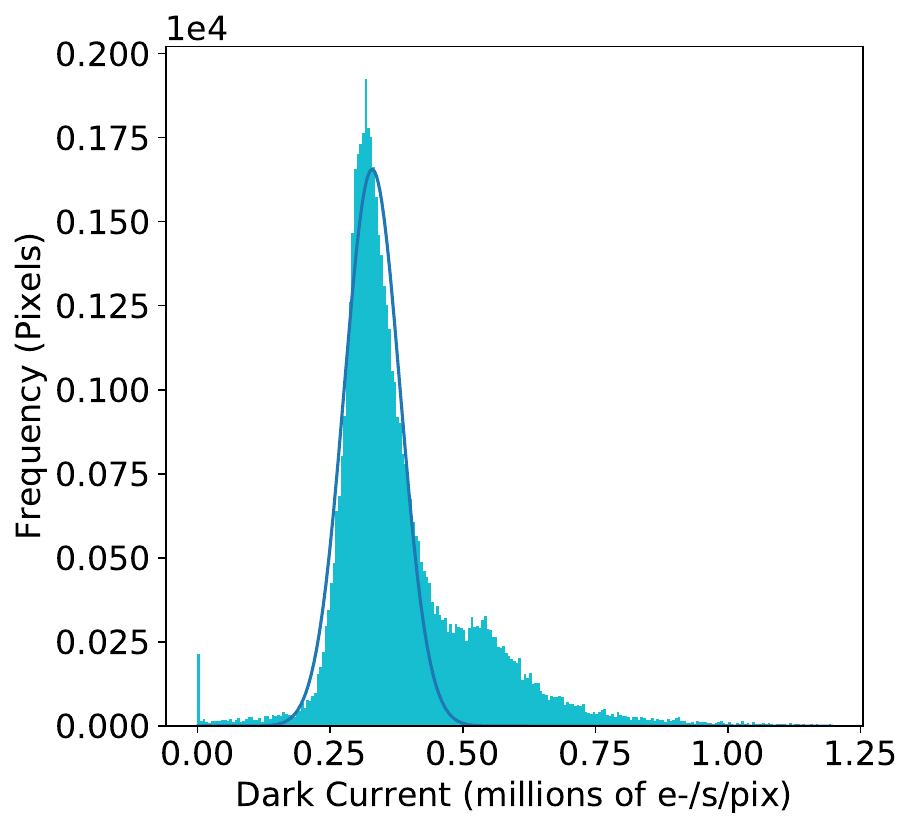}
    \caption{A histogram of individual pixel dark currents with SN23260 at 45 K within the cold masked edge of the device. This dark current is calculated via a subtraction of the 100\% and 0\% Tint 1 Hz data.}
    \label{fig:darkcurr}
\end{figure}

To estimate the dark current, we repeat this process with 100\% Tint frames subtracted by the 0\% Tint frames. We also multiply by the gain to convert into e-/s/pix. We perform the analysis on a region behind the cold mask and then across the unmasked area. First, we calculate the dark current in an "edge" region from (20:2028,10:40). We recover a Gaussian fit with mean of 330,000 e-/s/pix and sigma of 52,600 e-/s/pix, shown in Figure \ref{fig:darkcurr}. We then perform the analysis across the whole detector and with the bad pixel mask (which blocks the edge region among other locations). This region could receive ambient radiation originating from light leaks into the cryostat or reflected emission from the warm capacitors on the back of SN23260. We recover a Gaussian fit with a mean of 840,000 e-/s/pix and sigma of 131,000 e-/s/pix at 1 Hz.

For both the edge region and masked array, we recalculate dark current with 10 Hz data, recovering approximately 264,000 e-/s/pix and 777,000 e-/s/pix, respectively. We believe the discrepancy is because Tint 0\% is not truly an IWR bias image and may be subject to both differences between a) ITR and IWR offsets and b) effective dark current during the minimum exposures. As longer exposures will be less impacted by the latter problem, we report 330,000 e-/s/pix as the device's 45 K dark current.

\begin{figure}
    \centering
    \includegraphics[width=\textwidth]{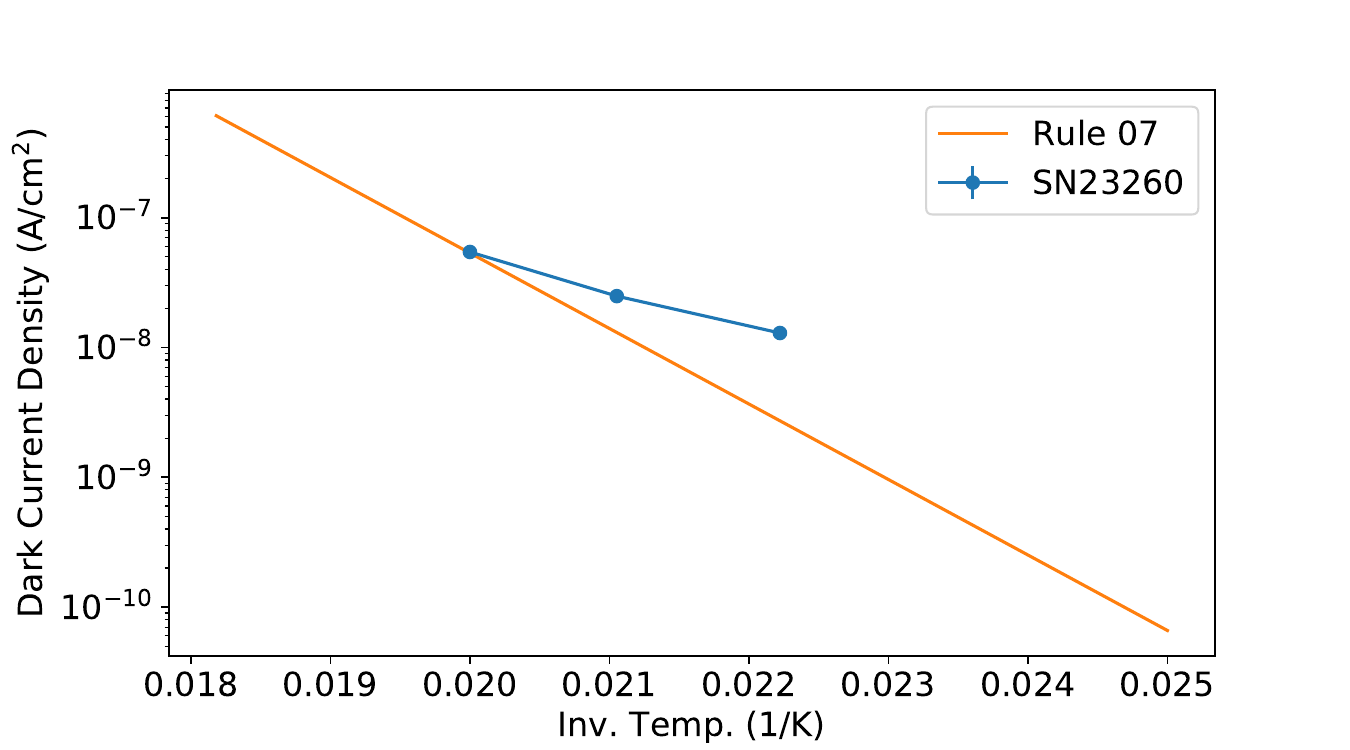}
    \caption{The cold masked edge dark currents for SN23260 at three temperatures versus the empirical prediction, Rule 07, as derived by Tennant et al. 2008 \cite{tennant2008JEMat..37.1406T}. Our device displays a shallower dark current slope than Rule 07 predicts which may be due to warm electronics glow providing a limiting dark current.}
    \label{fig:mux}
\end{figure}

We repeat the process in the edge region for 10 Hz dark data taken with the detector at 47.5 K and 50 K. We find a dark current at 47.5 K of 506,000 e-/s/pix and a dark current at 50 K of 1.107 million e-/s/pix. Tennant et al. determined an empirical expression for the dark current in MCT devices, referred to as "Rule 07" \cite{tennant2008JEMat..37.1406T}. We plot our dark currents for 45 K, 47.5 K, and 50 K in A/cm$^2$ alongside Rule 07 in Figure \ref{fig:mux}, scaled to the 50 K result. The slope of our dark current is shallower than predicted by Rule 07 which may indicate that SN23260 is hitting an electronics glow or ambient background limited floor as it approaches 45 K. However, more data are needed at lower temperatures to confirm.

\subsection{Linearity}
\label{subsec:linearity}

To assess linearity, we utilize the same data used in the gain calculations but work from a region C (850:1050, 500:700), located 400 pixels shortwards in x from Region A's center. This is done to place the zone of interest in a location that will saturate at 70\% Tint rather than 95\% Tint. We produce a mean image for each set of 128 frames and we then examine the behavior for individual pixels.

\begin{figure}
    \centering
    \includegraphics[width=\textwidth]{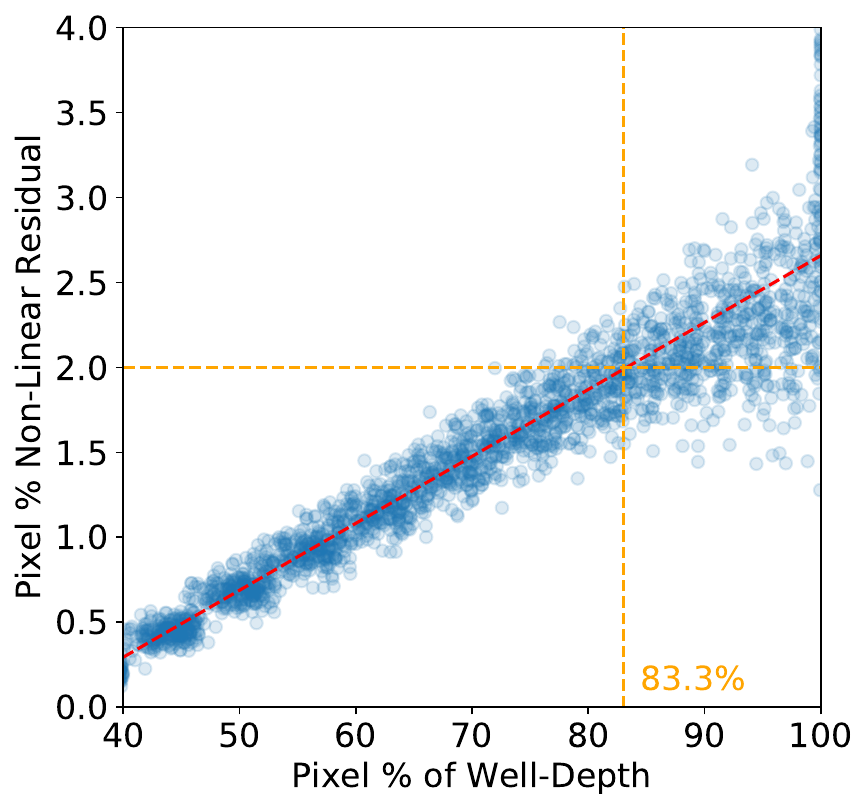}
    \caption{A scatterplot of non-linear residual versus well-depth for all pixels in region C over a series of integrations that culminate with saturation.}
    \label{fig:linearity}
\end{figure}

We first define a pixel's maximum as its value in the 100\% (saturated) frame and then quote its data as \% of full-well. We use the data points below 40\% to fit a linear trend to the data as signal level versus exposure time. We calculate the non-linear residual for each well-depth greater than 40\%. In Figure \ref{fig:linearity}, we present a scatterplot of non-linear residuals versus their corresponding well-depth. We fit a line to the non-linear residuals and find that most pixels experience $<$ 2\% non-linearity at $<$ 83\% well-depth. While not yet quantified, there is no reason to suspect variations in this behavior over time, suggesting it can be calibrated.

\subsection{1/f Noise}
GeoSnap suffers from 1/f noise, indicating that greater time between exposures result in more noise between frames. As described in Leisenring et al. 2023 \cite{jarron2023AN....34430103L}, it is more accurate to say that this noise scales with the number of frames. Thus, two images taken an equal number of frames apart will have the same 1/f noise, regardless of the frame rates at which they were taken. Leisenring et al. also reported that as the signal onto the detector increased, the 1/f noise slope decreased.

\begin{figure}
    \centering
    \includegraphics[width=\textwidth]{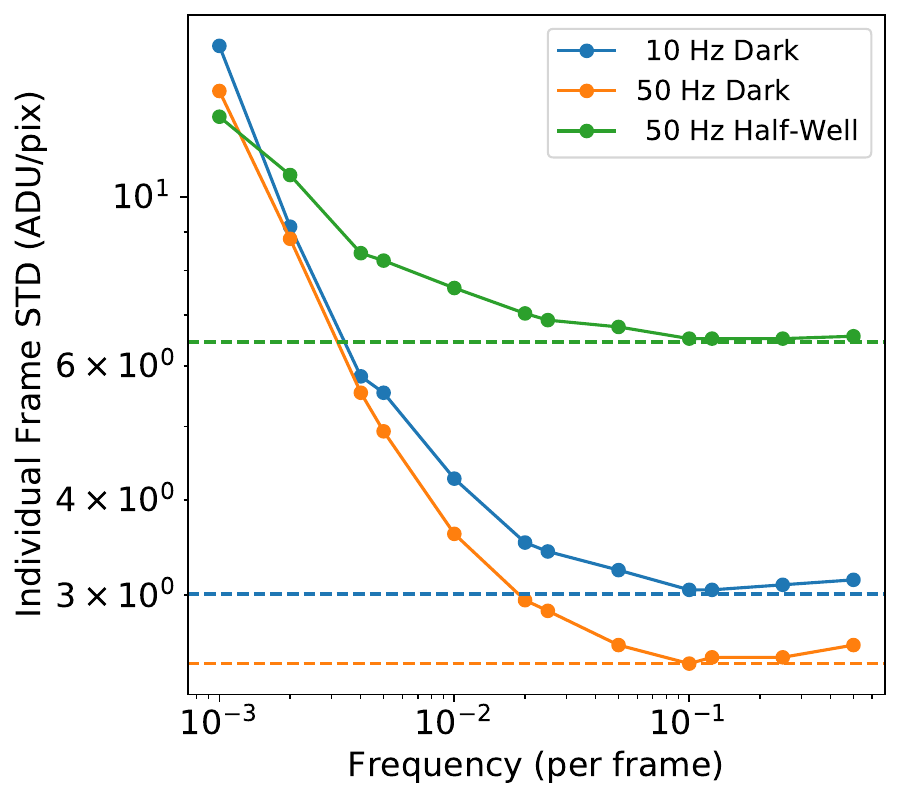}
    \caption{1/f behavior for SN23260 for different frame rates and well-depths. Dashed-lines are used to show the theoretical noise limited floor for each data set, assuming the only sources of noise in the individual frames are read noise and shot noise added in quadrature. The significance of 1/f noise is negligible if one modulates the signal at a high frequency (per frame).}
    \label{fig:one_over_f}
\end{figure}

We verify this 1/f behavior in SN23260 by taking 1,000 dark images at 10 Hz and 50 Hz. We also take an additional 1,000 half-well images at 10 Hz using the ammonia filter. Frames are subtracted from other frames in specific integer spacings to replicate sampling at all accessible frequencies and we work from region B. We display the data in terms of the "frequency (per frame)" in Figure \ref{fig:one_over_f}. Frequency (per frame) is the sampling frequency divided by the frame rate and is a means of showing different frame rate curves in terms of the same number of images between a subtraction. So 0.1 frequency (per frame) at 10 Hz means there is 1 second (10 images) in a chop cycle while at 50 Hz it is 0.2 seconds (10 images) in a chop cycle. In both cases, frame 1 would be subtracted by frame 6, 2 by 7, and so on until it repeats. Similar to the UM GeoSnap in MIRAC-5\cite{jarron2023AN....34430103L}, SN23260 reaches the shot noise limited floor with high frequency sampling. The behaviors of the 1/f are nearly identical when viewed in terms of number of frames though their floors differ due to the shot noise expected from the three different signal levels. The 10 Hz half-well data has a far greater shot floor and demonstrates that as the signal onto the detector increases, the 1/f noise slope decreases. In all cases, given a fixed level of shot noise, one can mitigate 1/f noise by sampling fast enough.

\subsection{Quantum Efficiency}
To verify the quantum efficiency of our detector, we can utilize a flux measurement taken at the detector plane. We can write the average signal per pixel ($\mu_{pix}$) in SN23260 as:

$$\mu_{pix} = \frac{f(\lambda)*QE(\lambda)*A_{pix}*t}{g*E_\gamma}$$

In the above, $f(\lambda)$ is the flux at the detector plane, $QE(\lambda)$ the quantum efficiency in a narrow bandpass, $A_{pix}$ the pixel area, $t$ the exposure time, $g$ the detector gain, and $E_\gamma$ the average photon energy in the bandpass. By working only with narrow bandpass filters such as the ammonia or 11 micron filter, we simplify our estimate as the anticipated blackbody emission, throughput, photon energy, and quantum efficiency should all be close to constant within the measurement regime.

To calculate quantum efficiency, we must solve this equation for $QE(\lambda)$. The only unknown variable is $f(\lambda)$, the flux at the detector plane. To determine $f(\lambda)$, we utilize a small single-cell photovoltaic (PV) MCT detector from TIS: J19D13. The J19D13 detector is well-calibrated with a known QE from 2 to 16 microns. J19D13 has a peak responsivity of 5.70 A/W with 10\% systematic uncertainty (via discussion with TIS). It operates at 77 K and is 1.0 mm by 1.0 mm square detector active region. We mount the J19D13 to the same position as SN23260 such that is should occupy a 55.55 by 55.55 pixel space centered at pixel (1024,1024). Thus, the pixel region 1000:1048 (exclusive) in both X and Y should experience the same net flux on it as the calibrator detector.

To obtain our throughput estimate, we measure the median pixel signal in the J19D13 region on SN23260 for several temperatures and filters, as shown in Table \ref{tab:qe_table}. We warm the cryostat, swap in J19D13, and then cool the cryostat again, leaving all other elements untouched between the thermal cycles. We repeat the same combination of tests. The voltage across J19D13 is measured via a voltmeter. The voltage is then converted to a current using the known total resistance of 10,259.7 Ohms. Then, this current is multiplied by the corresponding responsivity ($\pm$ 10\%) in the bandpass, according to the relative responsivity curve provided by TIS and the peak responsivity, to get a wattage onto the detector. Using the active area of the detector, we calculate the flux onto J19D13. We tested the uncertainty of our J19D13 voltages due to drifts in its bias value, stability of the source, and variations across cooldowns. We found that systematic uncertainty across cooldowns (of 3\%) dominated over all other sources of uncertainty, with the exception of very low flux measurements (such as a 200 K source measured at 11 micron filter measurements). Therefore, all J19D13 voltages are taken to have a minimum 3\% uncertainty.

For both SN23260 and J19D13, we take "ambient dark" data (i.e., with the source on but the filter wheel on the blank slot), and subtract that value from the image and measured voltage, respectively. This is necessary since even in the blank filter position the detector senses some light.

We provide the estimated flux (and corresponding throughputs) based on the J19D13 data in Table \ref{tab:qe_table}. For the SN23260, we based our choice of frame rate and Tint on remaining within the linear regime for the data set. We then calculate the necessary SN23260 QE to match the measurement pairs. Working with the 300 K data, we find that the ammonia (10.6 micron) filter results in a QE of 79.7 $\pm$ 8.3 \% while the 11 micron filter results in a QE of 79.0 $\pm$ 8.0 \%. For comparison, TIS estimated QEs of 88\% at 8.69 microns, 76\% at 10.77 microns, and 65\% at 12.02 microns.

\begin{table}[ht]
\caption{SN23260 and J19D13 warm source measurements for several filters and source temperatures. Net values are found for both detectors after "ambient" dark subtraction.} 
\label{tab:qe_table}
\begin{center}       
\begin{tabular}{|l|l|l|l|l|l|l|l|}
\hline
\rule[-1ex]{0pt}{3.5ex}  Source & Filter & Net &  Net & SN23260 & Measured  & Estimated Flux & QE\\
\rule[-1ex]{0pt}{3.5ex}  Temp. &  & J19D13 & J19D13 & Exp. & SN23260 & SN23260 & \\
\rule[-1ex]{0pt}{3.5ex}  &  & Signal &  Flux  & Time & Signal &  Signal & \\
\rule[-1ex]{0pt}{3.5ex} (K) &  & (mV) & (erg/cm$^2$/s) & (ms) & (e-/pix/frame) & ($\gamma$/pix/frame) & (\%) \\
\hline
\hline

\rule[-1ex]{0pt}{3.5ex} 200 & Amm. & 0.32 $\pm$ 0.02 & 5.41 $\pm$ 0.60 & 19.860 & 1.66E6 $\pm$ 9E3 & 1.86E6 $\pm$ 2E5 & 89.5 $\pm$ 9.9 \\
\hline
\rule[-1ex]{0pt}{3.5ex} 200 & 11 $\mu$m & 0.07 $\pm$ 0.01  & 1.19 $\pm$ 0.24 & 19.860 & 4.18E5 $\pm$ 2E3 & 4.35E5 $\pm$ 9E4 & 96.2 $\pm$ 19.4 \\
\hline
\rule[-1ex]{0pt}{3.5ex}  &  &   &  &  &  &  &  \\
\hline

\rule[-1ex]{0pt}{3.5ex} 250 & Amm. & 1.48 $\pm$ 0.02 & 24.8 $\pm$ 2.6 & 3.146 & 1.04E6 $\pm$ 5E3 & 1.35E6 $\pm$ 1E5 & 76.7 $\pm$ 8.0 \\
\hline
\rule[-1ex]{0pt}{3.5ex} 250 & 11 $\mu$m & 0.33 $\pm$ 0.02  & 5.4 $\pm$ 0.60 & 11.624 & 8.46E5 $\pm$ 4E3 & 1.15E6 $\pm$ 1E5 & 73.8 $\pm$ 8.7 \\
\hline
\rule[-1ex]{0pt}{3.5ex}  &  &  &  &  &  &  &  \\
\hline

\rule[-1ex]{0pt}{3.5ex} 300 & Amm. &  3.57 $\pm$ 0.10 & 60.0 $\pm$ 6.2 & 0.885 & 7.32E5 $\pm$ 4E3 & 9.18E5 $\pm$ 1E5 & 79.7 $\pm$ 8.3 \\
\hline
\rule[-1ex]{0pt}{3.5ex} 300 & 11 $\mu$m &  0.71 $\pm$ 0.01 & 11.7 $\pm$ 1.2 & 5.972 & 1.01E6 $\pm$ 5E3 & 1.28E6 $\pm$ 1E5 & 79.0 $\pm$ 8.0 \\
\hline

\end{tabular}
\end{center}
\end{table}

\section{Conclusion}
\label{sec:conclusion}

We analyzed the performance of the 2048x2048 SN23260 GeoSnap with a B0 ROIC. The proper quantification of this device is paramount as it is one of the candidate detectors for METIS on the ELT. We measured various properties of the detector, finding:
\begin{enumerate}
    \item At 10 Hz and in the inner 1848x1848 pixels, there are 3.17\% bad pixels, 0.7\% of which are contained within twenty-nine "leopard" spots and two "glow" spots on the detector.
    \item The detector has an average gain of 194.7 e-/ADU and mean well depth of 2.75 million e-/pix.
    \item The detector has an average read noise of 360 e-/pix.
    \item The detector has a dark current of approximately 330,000 e-/s/pix at 45 K in the cold masked region measured at 1 Hz frame rate.
    \item The detector achieves $<$ 2\% non-linearity at $<$ 83\% well-depth.
    \item The detector displays 1/f noise matching trends observed with the previously tested UM GeoSnap.
    \item The detector has a QE of 79.7 $\pm$ 8.3 \% at 10.6 microns. 
    
\end{enumerate}

SN23260's performance matched expectations based on our previous experience. In the future, we would like to further explore possible detector glow or ambient background. By obtaining information above 50 K and below 45 K, we can check how the device performs relative to Rule 07. Although persistence is not expected for CTIA devices, verification of this statement would be ideal. We will also verify the temporal stability of all detector parameters, especially linearity. Although the main testing for SN23260 is complete, more thorough tests for it and other METIS candidate GeoSnap will be completed in the future.

\acknowledgments 
 We thank Vincent Douence from TIS for his major contributions in tuning and operating of GeoSnap SN23260 and other equipment. We also thank Henry Yuan and David Bond from TIS for their assistance with planning and utilizing the J19D13 detector for the QE assessments. We are grateful to the University of Michigan for their support of the infrared detector test lab.
\bibliography{report} 
\bibliographystyle{spiebib} 

\end{document}